\documentclass[journal,onecolumn]{IEEEtran}
\usepackage{graphicx}

\usepackage{cite}

\ifCLASSINFOpdf
\else
\fi
\usepackage{amsmath}
\usepackage{algorithmic}

\usepackage{array}

\ifCLASSOPTIONcompsoc
 \usepackage[caption=false,font=normalsize,labelfont=sf,textfont=sf]{subfig}
\else
  \usepackage[caption=false,font=footnotesize]{subfig}
\fi

\usepackage{stfloats}
\ifCLASSOPTIONcaptionsoff
  \usepackage[nomarkers]{endfloat}
 \let\MYoriglatexcaption\caption
 \renewcommand{\caption}[2][\relax]{\MYoriglatexcaption[#2]{#2}}
\fi

\begin{document}
%
\title{High Permittivity Dielectric Field-Plated Vertical (001) $\beta$-Ga$_2$O$_3$ Schottky Barrier Diode with Surface Breakdown Electric Field of 5.45 MV/cm and BFOM of $>$ 1 GW/cm$^{2}$}
%
%
%

\author{Saurav~Roy,~
        Arkka~Bhattacharyya,~
        Praneeth~Ranga,~
        Heather~Splawn,~
        Jacob~Leach,~
        and~Sriram~Krishnamoorthy
\thanks{This work was supported by the National Science Foundation (NSF) under grant DMR-1931652 and by the Air Force Office of Scientific Research under award number FA9550-21-0078. Any opinions, findings, conclusions or recommendations expressed in this article are those of the author(s) and do not necessarily reflects the views of the united states air force.}
\thanks{S. Roy, A. Bhattacharyya, P. Ranga, S. Krishnamoorthy are with department of electrical and computer engineering, The university of Utah, Salt Lake City, UT 84112, USA (e-mail: saurav.roy@utah.edu, sriram.krishnamoorthy@utah.edu).}
\thanks{H. Splawn and J. Leach are with Kyma Technologies, Inc., Raleigh, NC 27617, USA}}

\maketitle

\begin{abstract}
This paper presents vertical (001) oriented $\beta$-Ga$_2$O$_3$ field plated (FP) schottky barrier diode (SBD) with a novel extreme permittivity dielectric field oxide. A thin drift layer of 1.7 $\mu m$ was used to enable a punch-through (PT) field profile and very low differential specific on- resistance (R$_{on-sp}$) of 0.32 m$\Omega$-cm$^{2}$. The extreme permittivity field plate oxide facilitated the lateral spread of the electric field profile beyond the field plate edge and enabled a breakdown voltage ($V_{br}$) of 687 V. The edge termination efficiency increases from 13.5 $\%$ for non-field plated structure to 63 $\%$ for high permittivity field plate structure. The surface breakdown electric field was extracted to be of 5.45 MV/cm at the center of the anode region using TCAD simulations. The high permittivity field plated SBD demonstrated a record high Baliga’s figure of merit (BFOM) of 1.47 GW/cm$^{2}$ showing the potential of Ga$_2$O$_3$ power devices for multi-kilovolt class applications.
\end{abstract}


%
\IEEEpeerreviewmaketitle

\section{Introduction}
%
%
%
%
\IEEEPARstart{T}{he} rise of $\beta$-Ga$_2$O$_3$ as the next generation power electronic material due to its large bandgap ($\sim$4.6 eV) and very high Baliga’s figure of merit (3400$_{Si}$) \cite{pearton2018review} has enabled researchers to demonstrate multi-kilovolt class devices regardless of its recent inception. The estimated critical electric field of $\beta$-Ga$_2$O$_3$ (8 MV/cm) is significantly higher than GaN (3.3 MV/cm) and SiC (2.4 MV/cm) holding promise for applications with very high voltage blocking capabilities. Various power devices including SBDs and transistors in both lateral and vertical geometry have been demonstrated with high breakdown voltages \cite{li2019field, zhou2019high, lu20201, hu2018field, hu2018lateral, lv2020demonstration, yang20182300v, li20182, xia2019metal, jian2020temperature, 8779668, joishi2018low, li2019single, wang2019high, li20181230}.

Despite its huge potential, $\beta$-Ga$_2$O$_3$ poses several challenges regarding the management of electric field at the device corners and edges, causing premature breakdown of devices. The lack of p-type doping, which is widely used in GaN and SiC devices for edge field management also poses a fundamental challenge in device design and complicated hetero-integration is needed for $\beta$-Ga$_2$O$_3$ \cite{roy2020design}.  Even with non p-type edge terminations, the average electric field for demonstrated $\beta$-Ga$_2$O$_3$ devices are far lower than the theoretical material limit. Hence, search of robust and efficient alternative edge termination techniques are necessary to rival the existing ultra-wide bandgap material-based devices.   

In this paper, we demonstrate a field plated vertical Schottky barrier diode with an extreme permittivity (BaTiO$_3$/SrTiO$_3$) \cite{kim2003strain} field plate oxide \cite{lee2020high} with punch through design. By achieving a high dielectric constant of $\sim$325, we have managed to increase the surface electric field at breakdown to 5.45 MV/cm. A breakdown voltage of 687 V in conjunction with a very low R$_{on-sp}$ (0.32 m$\Omega$-cm$^{2}$) has resulted in a record high BFOM of 1.47 GW/cm$^{2}$. TCAD simulation is also performed to estimate the field distribution and the breakdown location in the fabricated devices.

 




\begin{figure}[t]
\centering
\includegraphics[width=5in,height=11cm, keepaspectratio]{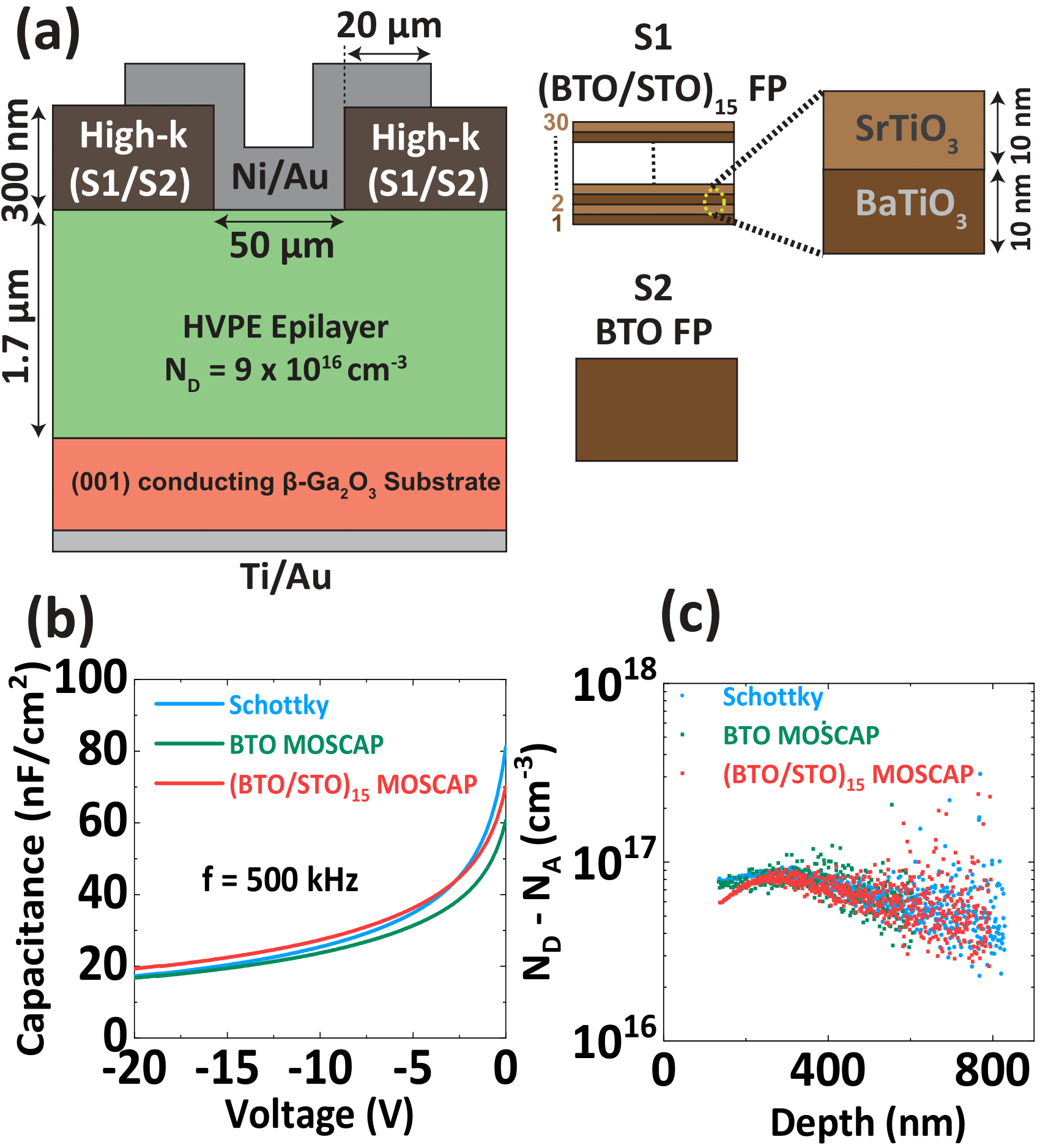}
\caption{(a) Schematic of Ga$_2$O$_3$ SBD with high-k dielectric field plate, where S1 is the (BTO/STO)$_{15}$ dielectric field oxide for (BTO/STO)$_{15}$ FP SBD and S2 is the BTO dielectric field oxide for BTO FP SBD. The expanded region of (a) shows the two BTO and STO layer thicknesses. (b) CV characteristics $\&$ (c) the extracted apparent charge density for the corresponding Schottky diode, BTO MOSCAP, and (BTO/STO)$_{15}$ MOSCAP.}
\label{fig1}
\vspace{-0.5cm}
\end{figure}

\section{Device Fabrication}
\label{sec2}
Device fabrication of the field plated SBD as shown in Fig. \ref{fig1}(a) started with the growth of the (001) $\beta$-Ga$_2$O$_3$ epilayers in a custom-designed subatmospheric vertical HVPE reactor using GaCl and O$_2$ precursors \cite{leach2019halide}. GaCl is produced in-situ by flowing Cl$_2$ over liquid Ga (7N purity), and is subsequently transported to the wafer surface through a quartz injector. Growth rates are controlled by the GaCl molar flow rates, and for these films the growth rates were $\sim$5 $\mu m$ per hour. Before the device fabrication, the samples were dipped in HF followed by HCl for 5 minutes. 15 layers of 10 nm thick BaTiO$_3$ and SrTiO$_3$ (BTO/STO)$_{15}$ (Total ~300 nm) were sputter deposited on the samples in oxygen ambient at room temperature. After the dielectric deposition, the samples were annealed at 700 $^0$C in oxygen ambient for 30 minutes to increase the dielectric constant. The sample was then patterned using standard photolithography and the active regions were opened-up using SF$_6$/Ar dry etching. About 50 nm of overetching is performed to ensure the complete removal of the dielectrics which was confirmed using transmission electron microscope imaging ($\sim$53 nm). Ni/Au (50/100 nm) metal stacks were then deposited on the sample using e-beam evaporation. Finally, Ti/Au (50/100 nm) ohmic contacts were sputter deposited on the back side of the sample. An additional sample with BaTiO$_3$ (BTO) dielectric as field plate oxide was also fabricated for comparison using the process flow described above. Besides the SBDs with (BTO/STO)$_{15}$ FP and BTO FP, SBDs with no FPs were also fabricated on a reference sample.

\section{Results And Discussions}
\label{sec3}
To understand effects of the dielectric deposition and the thermal annealing on the carrier concentration, capacitance voltage (CV) measurements were performed as shown in Fig. \ref{fig1}(b) for the schottky diode, BTO MOSCAP, and (BTO/STO)$_{15}$ MOSCAP. A doping concentration of 8-9$\times$10$^{16}$ cm$^{-3}$ is extracted from the CV plot as shown in Fig. \ref{fig1}(c). Dielectric constants of the deposited (BTO/STO)$_{15}$ and BTO layer were extracted to be $\sim$325 and $\sim$142 by depositing the dielectrics on an MOCVD grown (010) $\beta$-Ga$_2$O$_3$ with a delta doped layer as reported by Xia \emph{et al.} \cite{xia2019metal}.

\begin{figure}[t]
\centering
\includegraphics[width=5in,height=10cm, keepaspectratio]{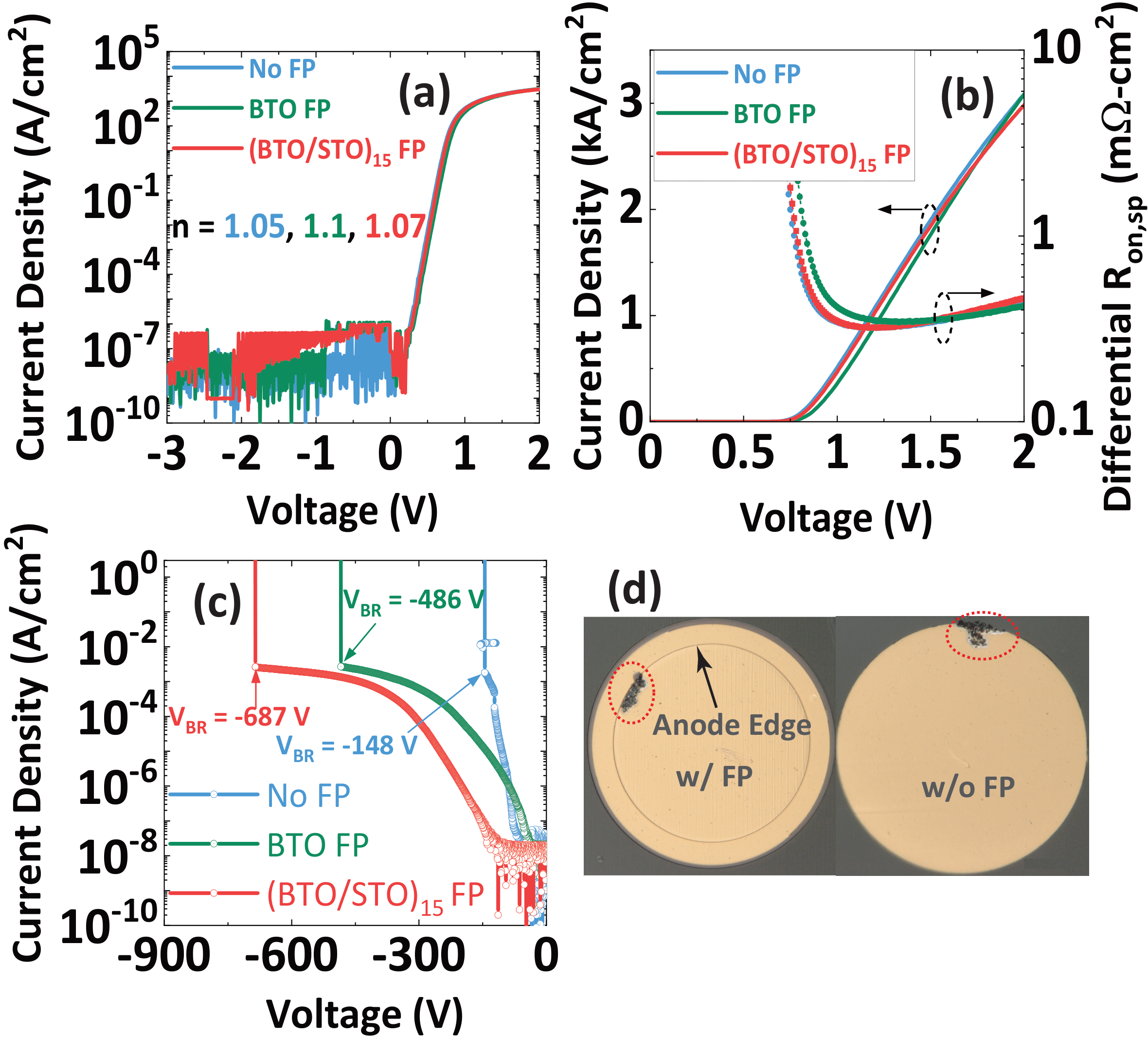}
\caption{(a) Log scale, $\&$ (b) linear scale IV characteristics of non-FP SBD, BTO FP SBD, and (BTO/STO)$_{15}$ FP SBD. (c) Breakdown characteristics of the three different SBD structures, and (d)  Microscope image of the fabricated device showing breakdown crater formation for the representative SBDs with and without FP.}
\label{fig2}
\vspace{-0.5cm}
\end{figure}

The typical current voltage plots in log and linear scale are shown in Fig. \ref{fig2}(a) and \ref{fig2}(b) respectively. The ideality factor (n) is extracted to be 1.04, 1.1 and 1.07 for the non- field plated SBD, BTO FP SBD, (BTO/STO)$_{15}$ FP SBD respectively. The R$_{on-sp}$ for the non-field plated, BTO FP, and (BTO/STO)$_{15}$ SBDs from the linear fit of the IV characteristics are extracted to be 0.33, 0.34, and 0.35 m$\Omega$-cm$^{2}$ respectively and the lowest differential R$_{on-sp}$ are 0.32, 0.34, and 0.32 m$\Omega$-cm$^{2}$ respectively. Nominally similar values of ideality factor and on-resistance measured in all the three samples indicates that the dielectric deposition and the dry etching process of the dielectric did not degrade the drift layer and the metal/semiconductor interface.

The breakdown characteristics for the SBDs with no FP, BTO FP, and (BTO/STO)$_{15}$ FP are shown in Fig. \ref{fig2}(c). The destructive breakdown voltages for the three devices are extracted to be 148 V, 486 V and 687 V respectively. The breakdown craters were observed at the anode edges as shown in the the optical microscope image of Fig. \ref{fig2}(d) for the FP and non-FP SBDs indicating the location of the breakdown. The increase in breakdown voltages for the high permittivity dielectric field plate oxide suggests the effectiveness of the edge termination and reduction in field crowding at the device edges due to the high permittivity of the FP dielectric.

\begin{figure}[t]
\centering
\includegraphics[width=5in,height=10cm, keepaspectratio]{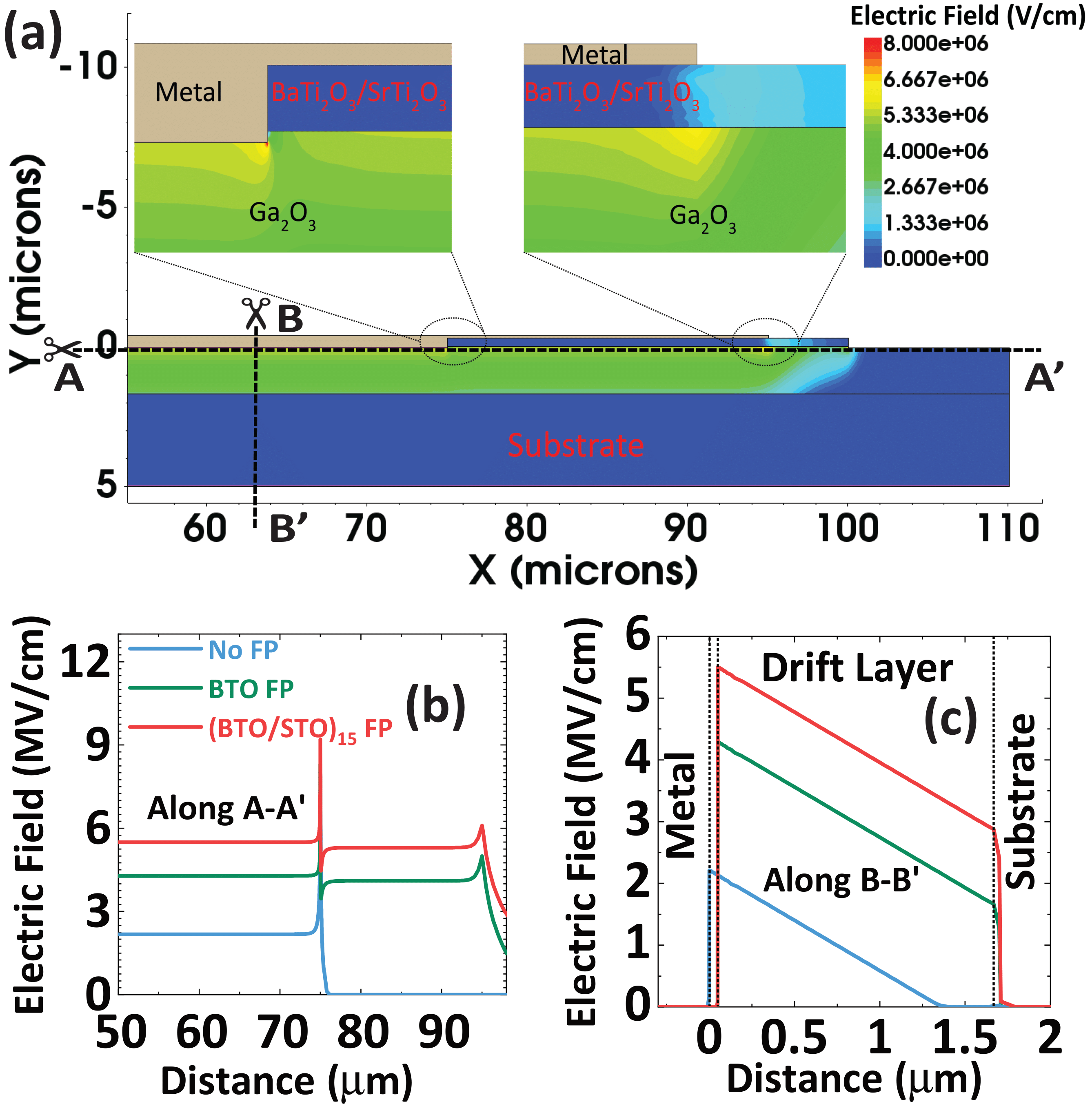}
\caption{(a) Electric field contour plot of the (BTO/STO)$_{15}$ FP SBD, (b) $\&$ (c) Simulated electric field profile for the three different SBDs for cutlines along the lateral and vertical directions respectively.}
\label{fig3}
\vspace{-0.35cm}
\end{figure}

\begin{figure}[t]
\centering
\includegraphics[width=5in,height=6cm, keepaspectratio]{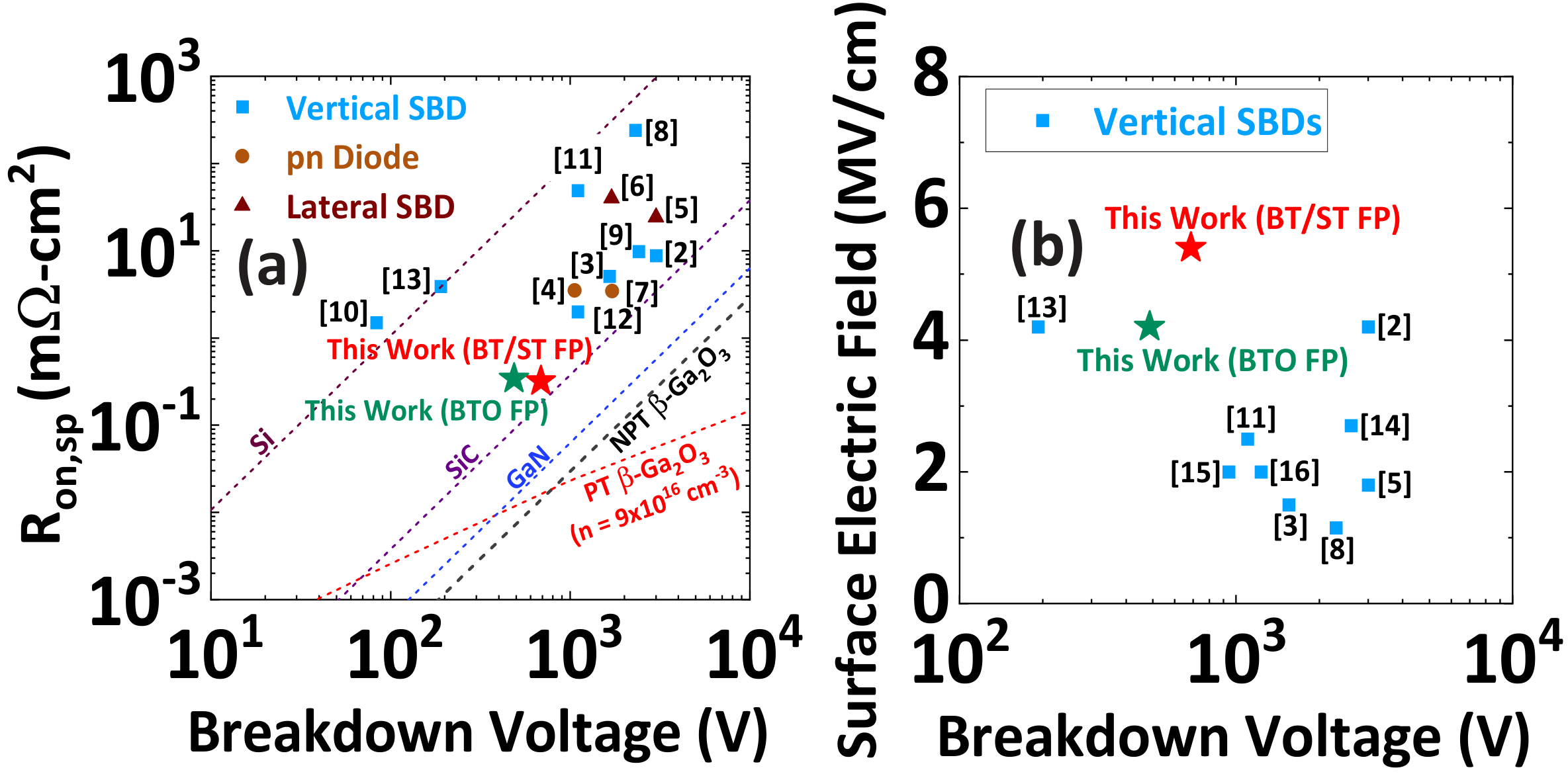}
\caption{Benchmark plot for (a) R$_{on-sp}$ vs $V_{br}$ and (b) Surface electric field vs $V_{br}$ comparing our device with other reported vertical and lateral diodes.}
\label{fig4}
\vspace{-0.5cm}
\end{figure}

The electric field simulations for the three SBD structures are also performed using Sentaurus TCAD software \cite{sentaurus} and results are plotted in Fig. \ref{fig3}(a), (b), and (c). Except for the non-field plated SBD, the other two devices reach punch through condition at breakdown as can be seen from Fig. \ref{fig3}(c). The calculated surface electric field (E$_{surf}$) for the non-punch-through and punch-through devices can be calculated as $E_{surf-npt} = \sqrt{\frac{2qN_DV_{br}}{\epsilon}}$ and $E_{surf-pt} =\frac{V_{br}}{t_{drift}} + \frac{qN_Dt_{drift}}{2\epsilon}$ respectively, where $q$ is the electron charge, $N_D$ is the donor concentration, $\epsilon$ is the relative permittivity, and $t_{drift}$ is the thickness of drift layer. Using the above two expressions, the surface electric fields can be calculated for SBD with no FP as 2.2 MV/cm, with BTO FP as 4.2 MV/cm, with (BTO/STO)$_{15}$ FP as 5.42 MV/cm. The calculated values for electric field match exactly with the simulation values as can be seen in Fig. \ref{fig3}(b), showing the validity of our simulations. This maximum surface electric field of 5.45 MV/cm for SBD with (BTO/STO)$_{15}$ FP is the highest value reported to date for all the Schottky barrier diodes. A peak electric field of 8-9 MV/cm was estimated at anode edges suggesting the location of the breakdown, thus, confirming the weak points for each type of device as seen in the microscope images shown in Fig. \ref{fig2}(d) . The edge termination efficiency ([Real $V_{br}$/Ideal $V_{br}$] ×100$\%$) increases from 44.2 $\%$ for the BTO FP to 63 $\%$ for (BTO/STO)$_{15}$ FP, where Ideal $V_{br}$ is calculated from the E$_{surf}$ expressions mentioned above by replacing E$_{surf}$ with E$_C$ (the critical electric field).

The extremely low R$_{on-sp}$ of 0.32 m$\Omega$-cm$^{2}$ (0.35 m$\Omega$-cm$^{2}$ from linear fit) and the breakdown voltage of 687 V for the (BTO/STO)$_{15}$ FP SBD results in a record high figure of merit of 1.47 GW/cm$^2$ (1.34 GW/cm$^2$ for the linear fit R$_{on-sp}$=0.35 m$\Omega$-cm$^{2}$). Fig. \ref{fig4}(a) shows the benchmark plots of R$_{on-sp}$ and breakdown voltage comparing performance of the (BTO/STO)$_{15}$ FP SBD with the existing literature reports. The performance metrics of our device is significantly better compared to the other reported devices. The surface breakdown electric field of 5.45 MV/cm is the highest electric field reported to date for $\beta$-Ga$_2$O$_3$ based SBDs as can be seen from Fig. \ref{fig4}(b). The device design of SBD demonstrated here using high permittivity field plate oxide has the potential to achieve the theoretical critical electric field of $\beta$-Ga$_2$O$_3$ showing promise for next generation power electronics.

\section{Conclusion}
\label{sec4}
In summary we have demonstrated a vertical (001) oriented $\beta$-Ga$_2$O$_3$ SBD with high permittivity dielectric field plate oxide showing extremely low R$_{on-sp}$ of 0.32 m$\Omega$-cm$^{2}$ and a high breakdown voltage of 687 V resulting in a BFOM value of 1.47 GW/cm$^2$ surpassing all the reported SBDs. The surface breakdown electric field of 5.45 MV/cm is the highest electric field reported to date for $\beta$-Ga$_2$O$_3$ SBDs. This demonstration indicates that the use of such high-permittivity dielectric field-plated structure enables high edge field termination efficiency and also enable studying the intrinsic critical field limits of ultra-wide band-gap semiconductors.


\bibliographystyle{IEEEtran}
\bibliography{ref}
\vspace{-0.5cm}
\end{document}